\documentclass[english,aps,preprint,superscriptaddress]{revtex4}
\usepackage[latin9]{inputenc}
\usepackage{array}
\usepackage{textcomp}
\usepackage{multirow}
\usepackage{amstext}
\usepackage{graphicx}
\usepackage{esint}
\usepackage{natbib}

\makeatletter

\usepackage{tikz}
\usetikzlibrary{external}

\makeatother

\usepackage{babel}
\begin{document}

\preprint{To be submitted to Phys. Rev. E}

\title{Permeability of porous foamy materials }

\author{V. Langlois}
\email{vincent.langlois@u-pem.fr}

\affiliation{Université Paris-Est, Laboratoire Géomatériaux et Environnement,
5 Boulevard Descartes 77454, Marne-la-Vallée Cedex 2, France}

\author{V.H. Trinh}
\email{van-hai.trinh@u-pem.fr}

\affiliation{Université Paris-Est, Laboratoire Modelisation et Simulation Multi
Echelle, UMR 8208 CNRS, 5 Boulevard Descartes 77454, Marne-la-Vallée
Cedex 2, France}

\author{C. Lusso}
\email{christelle.lusso@onera.fr}

\affiliation{Université Paris Est, Laboratoire Navier, UMR 8205 CNRS - École des
Ponts ParisTech - IFSTTAR cité Descartes, 2 allée Kepler, 77420 Champs-sur-Marne,
France}

\author{C. Perrot}
\email{camille.perrot@u-pem.fr}

\affiliation{Université Paris-Est, Laboratoire Modelisation et Simulation Multi
Echelle, UMR 8208 CNRS, 5 Boulevard Descartes 77454, Marne-la-Vallée
Cedex 2, France}

\author{X. Chateau}
\email{xavier.chateau@enpc.fr}

\affiliation{Université Paris Est, Laboratoire Navier, UMR 8205 CNRS - École des
Ponts ParisTech - IFSTTAR cité Descartes, 2 allée Kepler, 77420 Champs-sur-Marne,
France}

\author{Y. Khidas}
\email{yacine.khidas@u-pem.fr}

\affiliation{Université Paris Est, Laboratoire Navier, UMR 8205 CNRS - École des
Ponts ParisTech - IFSTTAR, 5 bd Descartes, 77454 Marne-la-Vallée Cedex
2, France}

\author{O. Pitois}
\email{olivier.pitois@ifsttar.fr}

\affiliation{Université Paris Est, Laboratoire Navier, UMR 8205 CNRS - École des
Ponts ParisTech - IFSTTAR cité Descartes, 2 allée Kepler, 77420 Champs-sur-Marne,
France}
\begin{abstract}
In this paper, we study the effects of both the amount of open cell
walls and their aperture sizes on solid foams permeability. FEM flow
simulations are performed at both pore and macroscopic scales. For
foams with fully interconnected pores, we obtain a robust power-law
relationship between permeability and membrane aperture size. This
result owns to the local pressure drop mechanism through the membrane
aperture as described by Sampson for fluid flow through a circular
orifice in a thin plate. Based on this local law, pore-network simulation
of simple flow is used and is shown to reproduce successfully FEM
results. This low computational cost method allowed to study in detail
the effects of the open wall amount on percolation, percolating porosity
and permeability. A model of effective permeability is proposed and
shows ability to reproduce the results of network simulations. Finally,
an experimental validation of the theoretical model on well controlled
solid foam is presented.
\end{abstract}
\maketitle

\section{introduction}

Foams are dispersions of gas in liquid or solid matrices. In liquid
foams, the structure of foams is made of membranes (liquid films separating
neighbor bubbles, also called walls), ligaments or Plateau's borders
(junction of three membranes) and vertex (junction of four ligaments).
Contrary to liquid foam, in which membranes are necessary to ensure
the stability of the foam, membranes can be partially or totally open
in solid foams. 

Permeability is a physical parameter that is used in many domains,
such as geophysics, soil mechanics, petroleum engineering, civil engineering
and acoustics of noise absorbing materials (e.g. polymeric foams).
As viscous dissipation is the most dissipative mechanism in the sound
propagation through porous materials, permeability (or flow resistivity)
is a key parameter governing the acoustical properties of such materials
\citep{johnson87a,allard09}. Different works have focused on the
effects of foams geometry on permeability: amount of closed walls
\citep{doutres13}, aperture of walls \citep{hoang12}, solid volume
fraction and ligament shapes \citep{kumar17,lusso17}. Authors deduce
some relations between permeability and studied parameters: solid
volume fraction, size of aperture,... These relations take into account
mechanisms acting at the scale of a bubble without taking into account
percolation. Indeed, in porous media, below a critical concentration of bonds between
pores inside a sample, the size of the interconnected porosity
is smaller than the sample height and no flow through sample is possible.
The classical Kozeny-Carman equation has to be modified to take into
account such percolation, e.g. in substituting the porosity by the
difference between the porosity and the critical porosity leading
to percolation \citep{mavko97}. Similary in foamy porous materials,
beyond a critical proportion of open walls, percolation has to occur.
Moreover, in the vicinity of the percolation, a small part of pores
is interconnected and the geometry of the pore network is complex.
To study more precisely the mesoscopic effect of the pore network
on permeability, numerical simulations should use large samples involving
a few hundred bubbles. However, as the size of samples increases,
the computational costs in FEM simulations become prohibitive. To
overcome similar difficulties in simulations of flow through porous
media, multi-scale approaches have been proposed \citep{kirkpatrick73,vogel00,vanmarcke10,jivkov13,xiong15,xiong16}:
at the scale of a throat between two linked pores, the relationship
between the flow rate passing through the throat linking pores and
the difference of pressure between pores is determined by numerical
simulations or analytical solutions (e.g. Hagen-Poiseuille equation);
at the macro-scale, pore-network simulations are performed to determine
the macroscopic permeability from local permeabilities found at the
local scale \citep{fatt56}. In this paper, we have attempted to use
a similar multi-scale approach to study the permeability of foamy
media.

Different numerical simulations are performed at different scales
to study the effect of aperture size and amount of closed walls. The
effect of the aperture size on partially open cell foam is studied
by using FEM simulations on periodic unit cells (PUC) involving the
Kelvin partition of space and containing two pores. To study the effect
of the proportion of closed walls, FEM simulations on larger samples
containing 256 pores are carried out in order to produce a flow at
the macroscopic scale and to simulate the complex flow through the
porous network. The mesoscopic effects induced by the structure of
the pore network are studied by pore-network simulations on large
(up to 2000 pores) lattice networks of interconnected pores interacting
via local permeabilities. A model of effective permeability based
on a calculation of the mean local permeability as in \citet{kirkpatrick73}
is used to describe the percolation threshold and the effect of mixing
local permeabilities. Finally, experimental measurements of permeability
performed on polymeric foams are compared to the model predictions.

\section{numerical simulations of foam permeability}

\subsection{FEM Simulations of fluid flow}

\subsubsection*{At the pore scale:}

\begin{figure}[h]
\centering
\includegraphics[scale=0.4]{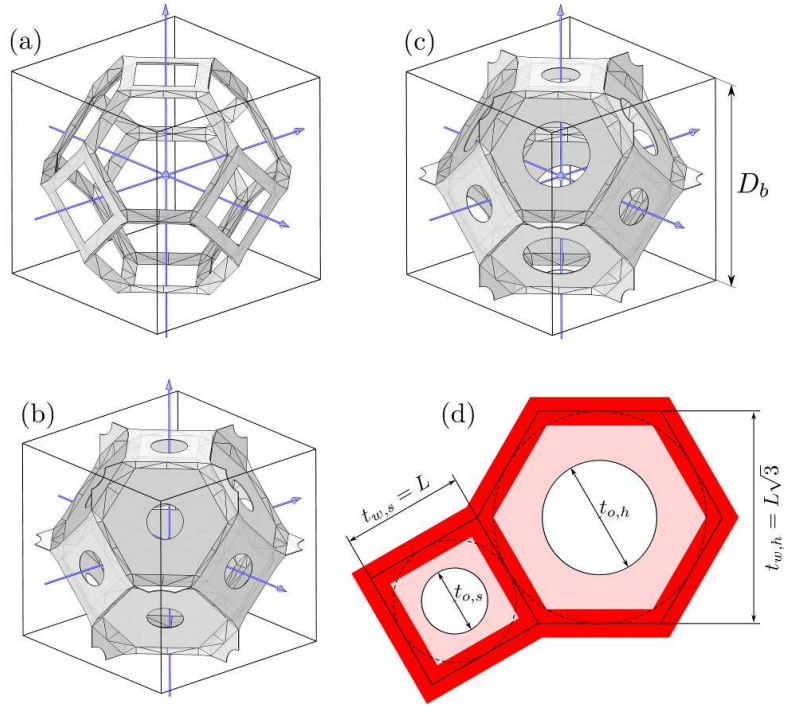}\caption{\label{fig:puc}PUC without walls (a), with identical aperture size
inside walls (b), with identical aperture rate (c), definitions of
the aperture size $t_{o}$ and the window size $t_{w}$ (d).}
\end{figure}

As shown on Fig. \ref{fig:puc}, a periodic unit cell containing two
pores of size $D_{b}$ is used to represent the pore structure in
foam samples \citep{perrot07}. The cell is based on the Kelvin paving
and is a 14-sided polyhedron corresponding to 8 hexagons and 6 squares.
The cell skeleton is made of idealized ligaments having length $L=D_{b}/(2\sqrt{2})$
and an equilateral triangular cross section of edge side $r=0.58D_{b}(1-\phi)^{0.521}$,
where $\phi$ is the gas volume fraction \citep{cantat13}. In the
reference configuration (Fig. \ref{fig:puc}a), the 14 cell windows
are fully open (i.e. without wall). As we are interested in the effect
of partial closure of the cell walls, we partially close the windows
by adding walls characterized with distinct circular aperture sizes.
Two kinds of simulations have been performed: (i) identical aperture
size on all windows (Fig. \ref{fig:puc}b), (ii) identical rate of
aperture $\delta_{ow}=t_{o}/t_{w}$ (Fig. \ref{fig:puc}c) where $t_{w}$
and $t_{o}$ are respectively the window size and the size of the
wall aperture as defined in Fig. \ref{fig:puc}d. The static viscous
permeability $K$ is computed from the solution of Stokes problem
\citep{adler92} for various porosity. The boundary value problem
is solved by using the finite element method (at convergence, the
mesh contains 214 412 tetrahedral elements) and the commercial software
COMSOL Multiphysics.

\subsubsection*{At the macroscopic scale:}

In order to study the flow properties on a larger scale, we have performed
numerical simulation for the flow of a Newtonian fluid through a periodic
network of Kelvins cells having a size $L\times L\times H=4\sqrt{2}\times4\sqrt{2}\times4$
in $D_{b}$ units (i.e. containing 256 pores), and a porosity$\phi$
equal to 0.9. Figure 2 shows an open cell foam sample made of 32 pores
(i.e. all the windows between adjacent cells are open). The macroscopic
intrinsic permeablity is computed from the averaging of the solution
of a Stokes problem set on the foam sample. In this study, the cell
windows are either closed or open with random spatial distribution
over the foam sample. For each value of the proportion of open windows,
the macroscopic intrinsic permeability is the average of numerical
simulations for 6 different samples. The resolution of the boundary
value problem is achieved through the Finite Element Method using
FreeFem++ Software. Typical discrete problem contains 1 400 000 Tetrahedra
and 8 000 000 degrees of freedom and is solved using a Message passing
Interface (MPI) on 4 processors.

\begin{figure}[h]
(a)\includegraphics[clip,scale=0.7]{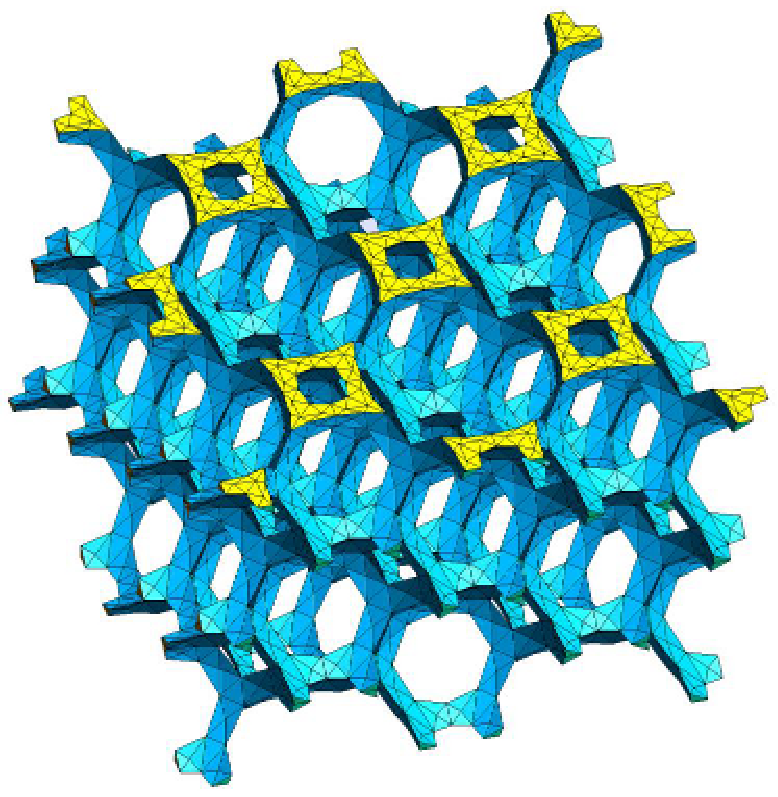}(b)\includegraphics[clip,scale=0.7]{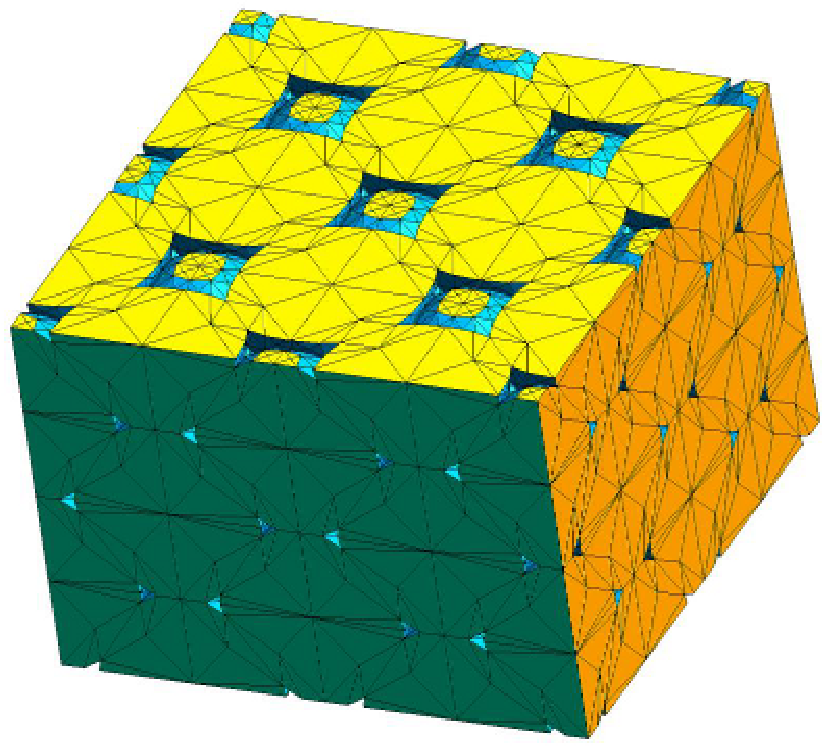}\caption{\label{fig:FEM2}FEM macro-scale samples: skeleton mesh (a) and porosity
mesh (b). For sake of visibility, a mesh of size $2\sqrt{2}\times2\sqrt{2}\times2$
($D_{b}$ units) is depicted in this figure.}
\end{figure}

\subsection{Pore-network simulations: }

\begin{figure}
\includegraphics[scale=0.75]{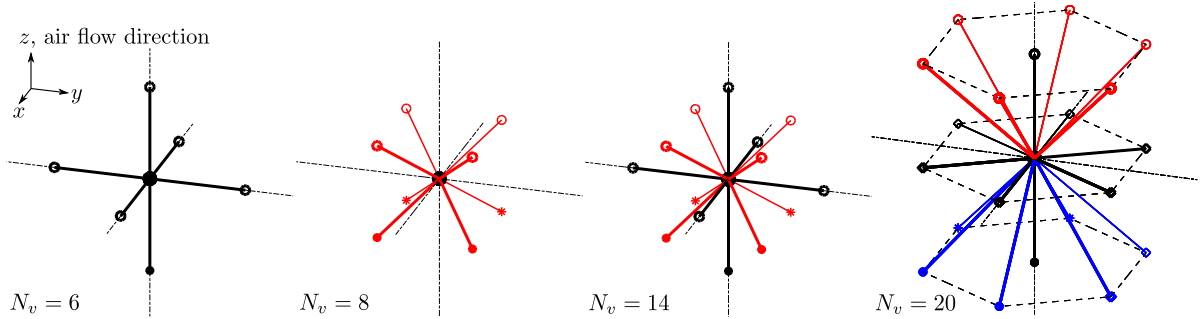}

\caption{\label{fig:Config}Network structures used in network simulations.}
\end{figure}

Effects of pore network features on permeability are studied on several
lattices of size $L\times L\times H=10\times10\times10$ ($D_{b}$
units) having different maximal numbers of neighbor pores $N_{v}$
(Fig. \ref{fig:Config}). In the case $N_{v}=14$, samples contain
2000 pores and the structure of pores corresponds to Kelvin's structure.
Boundary effects are avoided by resorting to periodic conditions imposed
in the directions perpendicular to the macroscopic flow. In this simple
model, we consider, for each pore, a unique value of pressure without
calculating the fluctuations of pressure and fluid velocity inside
the pore. At the local scale, the flow rate $q_{j\rightarrow i}$
from pore $j$ to pore $i$ is governed by the differential pressure
between the pores $\Delta P_{ij}=P_{j}-P_{i}$:

$q_{j\rightarrow i}=\frac{D_{b}}{\mu}k_{ij}\Delta P_{ij}$

where the coefficient $k_{ij}$ is the local permeability between
the pores $i$ and $j$.

At steady state and by considering incompressible fluid, the volume
of fluid inside pore $i$ is constant and the sum of flow rates coming
from neighbor bubbles is equal to zero, leading to: $\sum_{j=1}^{N_{v}}k_{ij}\left(P_{j}-P_{i}\right)=0$.
To generate a flow through the sample, a pressure difference is imposed
between top and bottom faces of the sample ($P_{top}=\text{\ensuremath{\Delta}}P_{sp}$,
$P_{bot}=0$). By considering these boundary conditions, this previous
equation can take a matrix form:

\begin{equation}
\underline{K}\left[P_{i}\right]=\left[S_{i}\right]\label{eq:matrix}
\end{equation}

where $\left[P_{i}\right]$ is a vector containing the pressure of
inner pores (pores located on top and bottom faces are excluded);
$\underline{K}$ is the matrix defined from local permeabilities ($-\sum k_{ij}$
along diagonal and $k_{ij}$ elsewhere) and $\left[S_{i}\right]$
is a vector containing zeros except for inner pores having top pores
as neighbors where $S_{i}=-\sum_{j_{top}}k_{ij_{top}}\text{\ensuremath{\Delta}}P_{sp}$.

As soon as the pore network links top to bottom and by considering
only the interconnected pores, $\underline{K}$ can be inverted and
the fluid pressure in each pore can be calculated from Eq.\ref{eq:matrix}.
Therefore, the macroscopic flow $Q$ and the macro permeability $K$
can be calculated as follows:

$Q=\sum_{i_{bot}}\sum_{j_{v}}q_{j_{v}\rightarrow i_{bot}}=\frac{D_{b}}{\mu}\sum_{i_{bot},j_{vi}}k_{i_{bot}j_{vi}}\Delta P_{i_{bot},j_{vi}}$ 

$K=\mu QH/L^{2}\text{\ensuremath{\Delta}}P_{sp}$

Different materials having different kinds of local permeability distribution
have been studied: two local permeabilities (binary mixture), a local
permeability mixed with zero permeability (closed walls) and two local
permeabilities mixed with closed walls. For each kind of local permeability
distribution, calculations are repeated from 200 to 400 times on different
random draws in order to calculate an average. For each random draw,
local permeabilities are randomly distributed over the network.

\subsection{Results and discussion}

\subsubsection*{Effect of the aperture size}

\begin{figure}
(a)\includegraphics[scale=0.8]{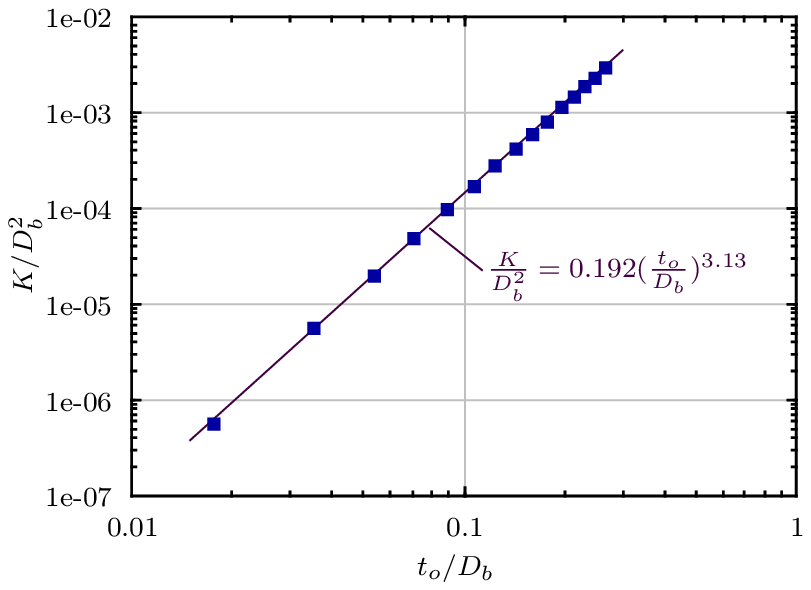}(b)\includegraphics[scale=0.8]{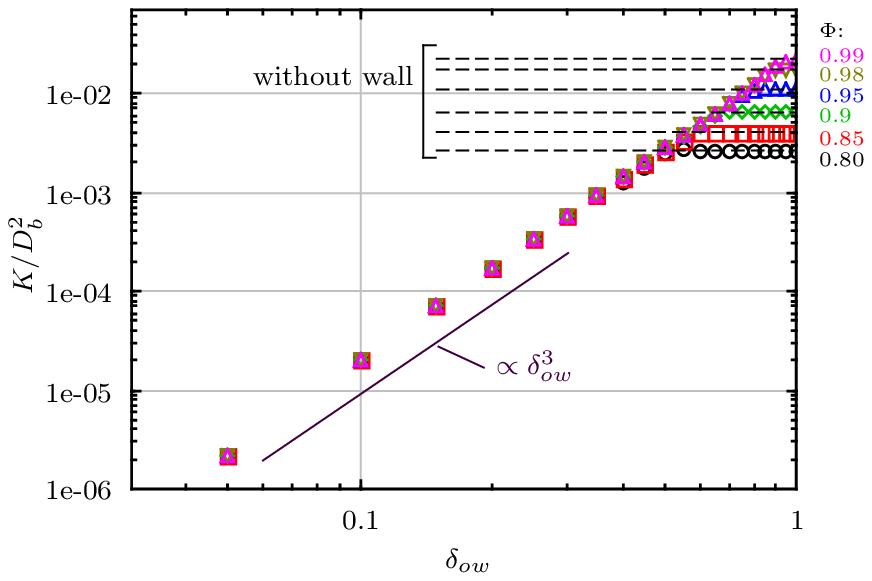}\\
(c)\includegraphics[scale=0.8]{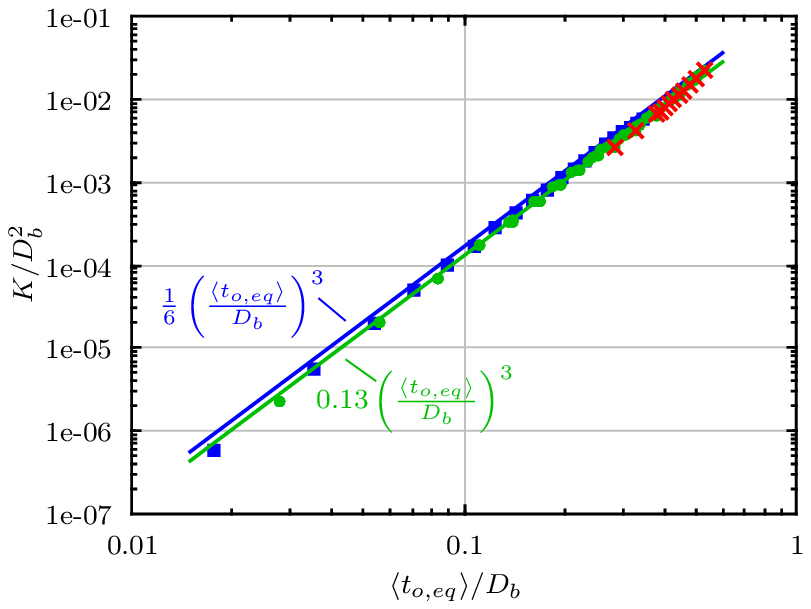}

\caption{\label{fig:local_permea}(a) FEM results at identical aperture size
with $\phi=0.98$, (b) FEM results at identical aperture rate for
various $\phi$, (c) Permeability as a function of the mean wall apertures:
FEM results (blue dot for identical aperture, green dot for identical
rate, red cross for ``no wall'' foam with $\phi$ varying from 0.8
to 0.99), network simulation with Sampson local permeabilities and
$N_{v}=14$ (blue line for identical aperture, green line for identical
rate). Note that the mean wall aperture is calculated without including
the four square windows which are parallel to the macroscopic flow
direction $\left\langle t_{o,eq}\right\rangle /D_{b}=(2t_{o,sq}+8t_{o,hex})/10D_{b}.$}
\end{figure}

FEM simulations on PUC at the pore scale for various aperture sizes
reveal a power-law relationship between permeability and aperture
size (Fig.\ref{fig:local_permea}a). Similarly the numerical results
for the dimensionless permeability of porous materials with same aperture
rate are well fitted by a power law when plotted in a ($\delta_{ow}$,
$K/D_{b}^{2}$) diagram (Fig.\ref{fig:local_permea}b). Note that,
for high aperture rates, the condition of identical aperture rate
is not observed due to the fact that the apertures should overlap
the ligaments, which is not allowed in our calculations. This artifact
leads to an artificial permeability plateau corresponding to the ``no
wall'' permeability. Apart from this artifact, FEM results show that
relationship between permeability and mean wall aperture is almost
unaffected by the porosity (i.e. the width of ligaments).

This power-law relationship is in agreement with a local interpretation
based on the pressure drop of the fluid passing through the wall aperture.
Indeed, Sampson \citep{sampson1891} solves analytically the problem
of the pressure drop $\Delta P$ occurring for an incompressible fluid
flow passing through a circular hole of diameter $d_{o}$ in a thin
plate:

\begin{equation}
\frac{Q}{\Delta P}=\frac{d_{o}^{3}}{24\mu}\label{eq:sampson}
\end{equation}

where $Q$ is the volume fluid flow rate passing through the hole
and $\mu$ is the fluid dynamic viscosity.

This relation arises from the fact that, at low Reynolds number, the
coefficient of fluid resistance $\zeta=2\Delta P/(\rho V_{o}^{2})$
is in general, proportional to the inverse of Reynolds number $R_{e}=V_{o}d_{o}\rho/\mu$
\citep{idel96}, where $V_{o}$ is the mean stream velocity in the
narrowest section of the orifice ($V_{o}=4Q/\pi d_{o}^{2}$).

After \citep{melick13}, the pressure drop through a hole of circular
shape is very close to the one obtained with a hole of squared shape
having the same area. We can deduce that the Sampson formula can be
extended to squared and hexagonal shape of aperture by taken into
account an equivalent diameter $t_{o,eq}$ defined from the surface
area of the aperture $S_{o}$: $t_{o,eq}=2(S_{o}/\pi)^{0.5}$. By
using such a definition for the aperture size and calculating a window
average of the aperture size, we can plot all FEM results on a same
graph. Fig. \ref{fig:local_permea}c shows that all data, including
the ones obtained without wall, follow the same trend. Therefore,
due to the peculiar pore geometry of foams, the pressure drop inside
such porous materials is governed by a local mechanism which is not
described by the usual Hagen-Poiseuille equation as it is done in
classical porous media \citep{fatt56,vogel00,jivkov13,xiong15}.

To check the ability of pore-network model to predict the permeability,
network calculations have been performed using local permeabilities
given by a Sampson equation:

\begin{equation}
k=t_{0}^{3}/24D_{b}\label{eq:local_sampson}
\end{equation}

In such simple simulated configurations (i.e. identical aperture size
or identical aperture rate), the network problem exposed in the previous
section can be solved analytically. Therefore, macroscopic permeability
is given by $K=2k_{sq}+2k_{hex}$, leading to $\frac{K}{D_{b}^{2}}=\frac{1}{6}\left(\frac{t_{0}}{D_{b}}\right)^{3}$
for identical aperture size and $\frac{K}{D_{b}^{2}}=\frac{1+3^{1.5}}{12}\left(\frac{5}{1+48^{0.5}}\frac{\left\langle t_{o,eq}\right\rangle }{D_{b}}\right)^{3}\approx0.13\left(\frac{\left\langle t_{o,eq}\right\rangle }{D_{b}}\right)^{3}$
for identical aperture rate. Fig. \ref{fig:local_permea}c shows that
network simulation results compare very well to FEM results. This
good agreement supports both the interpretation of the permeability
by using local permeabilities and the relevance of pore-network simulations.

\subsubsection*{Effect of closed walls}

\begin{figure}[h]
\includegraphics[scale=1.1]{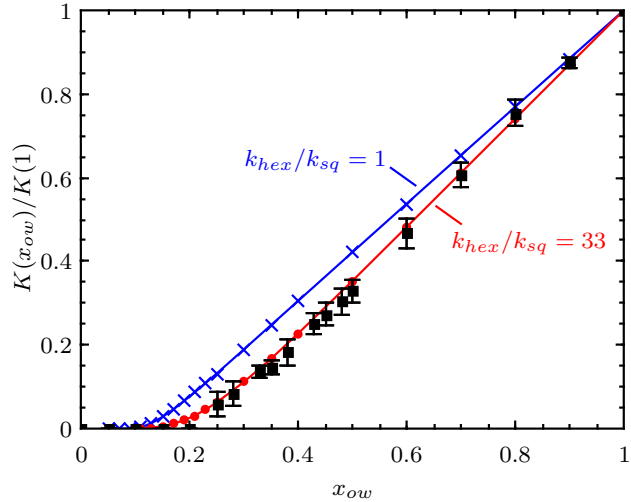}

\caption{\label{fig:mix_FEM}Dimensioless permeability $K(x_{ow})/K(1)$ as
a function of open walls fraction $x_{ow}$ for FEM simulations (black
square) and network simulations on samples mixing two local permeabilities
with various ratios $k_{hex}/k_{sq}$ and having a Kelvin structure
($N_{v}=14$). }
\end{figure}

Fig. \ref{fig:mix_FEM} shows the permeabilities calculated by FEM
simulations on large samples having random positions of closed walls
and various open walls fractions $x_{ow}$ (defined as the number
of open wall over the total number of walls in the porous sample).
For $x_{ow}>0.3$, permeability exhibits a quasi-affine dependence
on the open walls fraction $x_{ow}$. Below a critical concentration
$x_{ow}<0.2$, the fluid flow vanishes.

Network simulations have been performed by considering two local permeabilities,
$k_{hex}$ and $k_{sq}$, given by Sampson equation and associated
to squared and hexagonal windows as in Kelvin's structure ($N_{v}=14$).
For $\phi=0.9$, the hexagonal/square aperture ratio in FEM simulations
is close to $3.2$. The ratio between local permeabilities is therefore
close to $k_{hex}/k_{sq}=33(\approx3.2^{3})$. As shown in Fig. \ref{fig:mix_FEM}
and considering the margin of error, network simulations and FEM simulations
lead to the same results. Moreover, network simulations reveal that
the slope of the affine part of the function $K(x_{ow})$ depends
on ratio between local permeabilities.

\begin{figure}[h]
\includegraphics[scale=0.9]{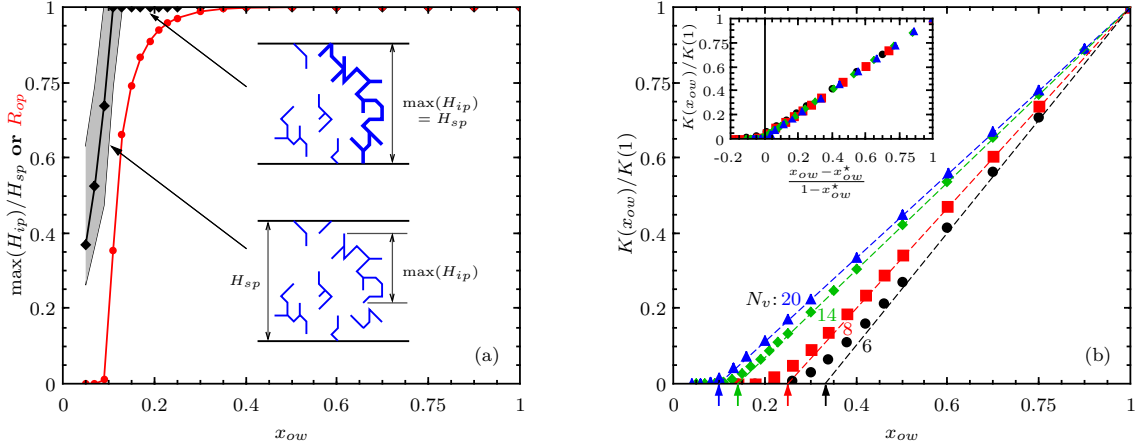}

\caption{\label{fig:permea_perco}Network simulations: (a) Maximal height of
the interconnected porosity (black line: median values - grey band:
1st and 99th percentile) and interconnected porosity (red line: mean
values) as a function of the open walls fraction $x_{ow}$ for $N_{v}=14$,
(b) Dimensioless permeability $K(x_{ow})/K(1)$ as a function of open
walls fraction $x_{ow}$ for various neighbor bubbles number $N_{v}$
(arrows point to the abscissa $x_{ow}=x_{ow}^{\star}$), inset: same
data with another abscissa $(x_{ow}-x_{ow}^{\star})/(1-x_{ow}^{\star})$.}
\end{figure}

Network simulations performed on different structures (Fig. \ref{fig:Config})
are helpful to study in details percolation effects for such foam
structures and to calculate both the heights of interconnected pores
$H_{ip}$ and the fraction of percolating porosity $R_{op}$ (Fig.
\ref{fig:permea_perco}a). In the case $N_{v}=14$, the maximal height
of interconnected pores is equal in average to the sample height $H_{sp}$
for $x_{ow}>0.1$, and connected porosity percolates throughout the
sample. Regarding the permeability (Fig. \ref{fig:permea_perco}b),
simulations performed with homogeneous local permeabilities show that
the slope of the affine part of $K(x_{ow})$ depends on the number
of neighbor bubbles $N_{v}$. The affine part of $K(x_{ow})$ intercepts
the abscissa to a critical concentration given by $x_{ow}^{\star}=2/N_{v}$.
Inset of Fig. \ref{fig:permea_perco}b shows that the ratio $K\left(x_{ow}\right)/K\left(1\right)$
in porous material having homogeneous local permeability is linearly
dependent on a single parameter $(x_{ow}-x_{ow}^{\star})/(1-x_{ow}^{\star})$
except in the vicinity of percolation.

\begin{figure}
\includegraphics[scale=1.1]{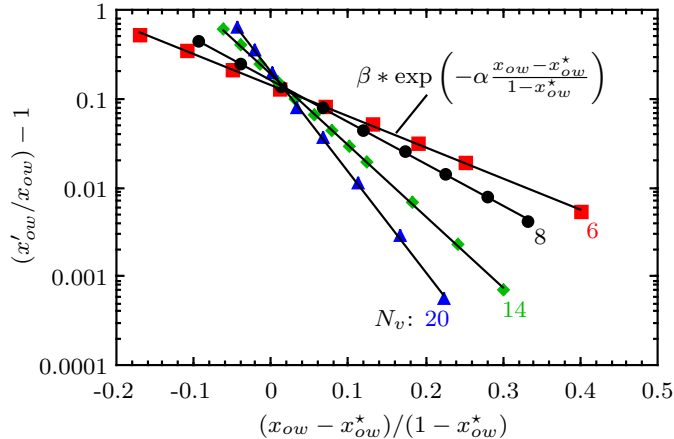}

\caption{\label{fig:Nm_pore}Excess fraction of open walls within the percolating
porosity as a function of the reduced fraction of open walls for various
neighbor numbers $N_{v}$. }
\end{figure}

A deeper analysis of results shows that percolation occurs on average
when $x_{ow}$ is in the range $\left[0.55x_{ow}^{\star};0.65x_{ow}^{\star}\right]$,
and that the fraction of open walls within the percolating porosity
$x_{ow}^{\prime}$ is larger than the global value $x_{ow}$. As shown
in Fig. \ref{fig:Nm_pore}, the relative gap between both fractions
of open walls is exponentially dependent on the reduced fraction $(x_{ow}-x_{ow}^{\star})/(1-x_{ow}^{\star})$.
The fraction $x_{ow}^{\prime}$ and the fraction of percolating porosity
$R_{op}$ can be approximated by the following equations:

\begin{equation}
\frac{x_{ow}^{\prime}}{x_{ow}}=1+\beta\exp\left(-\alpha\frac{x_{ow}-x_{ow}^{\star}}{1-x_{ow}^{\star}}\right)\label{eq:x_op}
\end{equation}

\begin{equation}
R_{op}=1-\min\left(1,\exp\left(-1.25-\alpha\frac{x_{ow}-x_{ow}^{\star}}{1-x_{ow}^{\star}}\right)\right)\label{eq:Rop}
\end{equation}

with $\alpha=2.56\left(\frac{N_{v}}{2}-1\right)+3$ and $\beta=0.123\left(\frac{N_{v}}{2}-1\right)^{0.23}$

From a practical point of view, these previous formulas might be useful
to estimate the percolating porosity and the open walls fraction inside
the percolating porosity by measuring for a real foam, the fraction
of open walls and the number of neighbor pores.

\section{Effective medium model for permeability}

\subsection{Description}

In this section, we present an effective permeability model of pores
network connected by local permeabilities ${k_{i}}$. This model is
based on a self-consistent calculation of the mean local permeability
and a calculation of the macroscopic permeability. Details leading
to Eqs. \ref{eq:mean_local_perm}-\ref{eq:effec_perm} are given in
Appendix.

The mean local permeability $\bar{k}$ is calculated iteratively from
\citep{kirkpatrick73,adler92}:

\begin{equation}
\frac{1}{\overline{k}+n\overline{k}}=\sum_{i}\frac{x_{i}}{k_{i}+n\overline{k}}\label{eq:mean_local_perm}
\end{equation}

with $x_{i}$ the fraction of local permeability $k_{i}$ and $n=\frac{N_{v}}{2}-1$.

In a few simple cases, this equation possesses analytical solutions
(e.g. binary mixture of local permeabilities, see Appendix).

The macroscopic effective permeability is then deduced from the mean
local permeability $\bar{k}$,

\begin{equation}
K\approx\frac{n}{2}\overline{k}\label{eq:effec_perm}
\end{equation}

In the next section, the present model is referred as EM model.

\subsection{Comparison between network simulations and EM model predictions}

In this part, we compare the predictions of EM model to the network
simulations. We consider successively three cases: (i) mixing of two
local permeabilities in a fully open material, (ii) mixing of closed
and open walls characterized by a single local permeability, (iii)
mixing of closed walls and two local permeabilities.

\begin{figure}[h]
\includegraphics[scale=1.1]{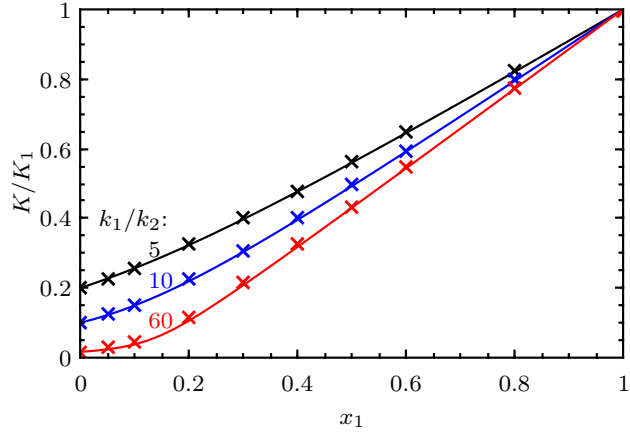}

\caption{\label{fig:mix_model1}Comparison of self-consistent model predictions
(full line) to network simulations (cross) with $N_{v}=14$. }
\end{figure}

(i) As shown in Fig. \ref{fig:mix_model1}, EM model predictions are
in good agreement with the network calculation results for various
ratios $k_{1}/k_{2}$. 

\begin{figure}[h]
\includegraphics[scale=1.1]{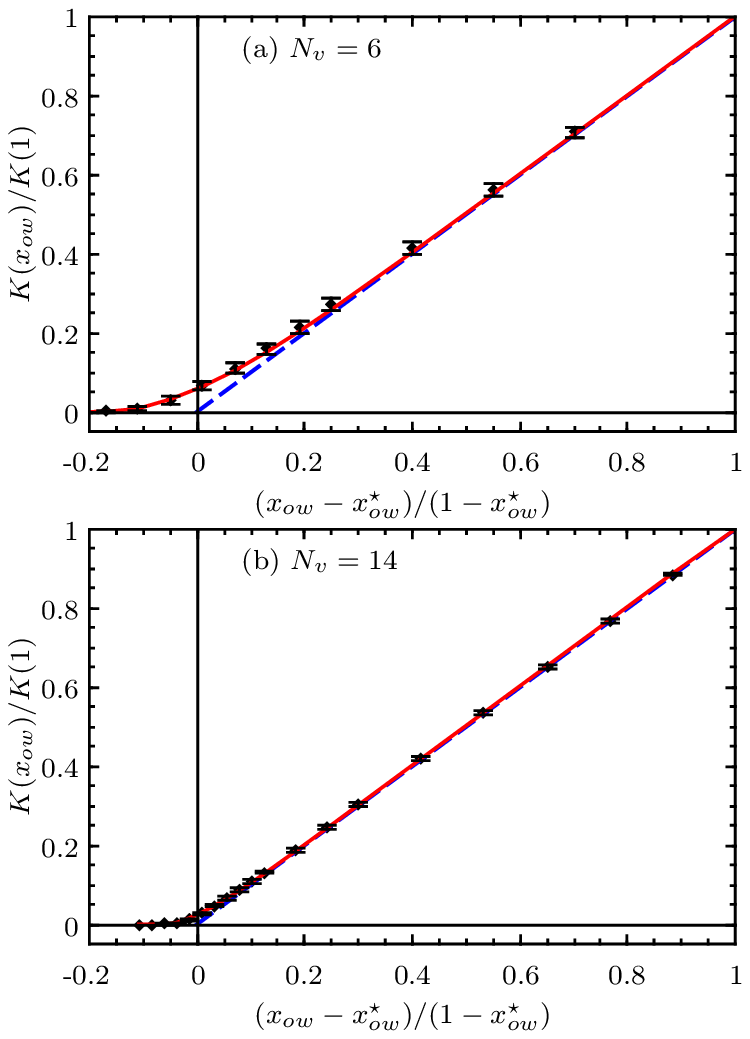}

\caption{\label{fig:permea_perco_model}Comparison between EM model and network
simulations (black diamond) for various neighbor pores numbers. ``EM0''
(dashed blue line) is based on the global open walls fraction (Eq.
(\ref{eq:5})), and ``EM1'' (red line) is based on the open walls
fraction within the percolating porosity (Eq. (\ref{eq:6})). }
\end{figure}

(ii) Mixing closed walls with identicaly open walls is a peculiar
case of the latter configuration: the local permeability associated
to the close walls is equal to zero. In this case, Eqs. \ref{eq:mean_local_perm}
and \ref{eq:effec_perm} have an analytical solution leading to:

\begin{equation}
\frac{K}{K_{1}}=\frac{x_{1}-x_{ow}^{\star}}{1-x_{ow}^{\star}}\label{eq:5}
\end{equation}

As shown in Fig. \ref{fig:permea_perco_model}, this solution ``EM0''
reproduces correctly the linear relationship between the permeability
and the parameter $(x_{ow}-x_{ow}^{\star})/(1-x_{ow}^{\star})$. However,
the permeability evolution is not accurately reproduced in the vicinity
of the percolation threshold. This discrepancy can be significantly
reduced if the fraction of open walls inside the percolating porosity
is used instead of the global open walls fraction (``EM1'' in fig.\ref{fig:permea_perco_model}),
leading to the following equation:

\begin{equation}
\frac{K}{K_{1}}=\frac{x_{1}^{\prime}-x_{ow}^{\star}}{1-x_{ow}^{\star}}R_{op}\label{eq:6}
\end{equation}

The fraction of open walls inside the percolating porosity can be
estimated from the global open walls fraction by using Eq. \ref{eq:x_op}.

For generalization purpose, one can write:

\begin{equation}
\frac{1}{\overline{k^{\prime}}+n\overline{k^{\prime}}}=\sum_{i}\frac{x_{i}^{\prime}}{k_{i}+n\overline{k^{\prime}}}\label{eq:mean_local_perm1}
\end{equation}

\begin{equation}
K\approx R_{op}\frac{n}{2}\overline{k^{\prime}}\label{eq:effec_perm1}
\end{equation}

where $x_{i}^{\prime}$ is the fraction of walls inside the percolating
porosity having a local permeability equal to $k_{i}$.

After Eq.\ref{eq:6}, the physical meaning of the critical concentration
$x_{ow}^{\star}=2/N_{v}$ is now clarify: at least two open walls
per bubble located in the percolating porosity are required to start
a sufficient interconnection of pores.

\begin{figure}[h]
\includegraphics[scale=1.1]{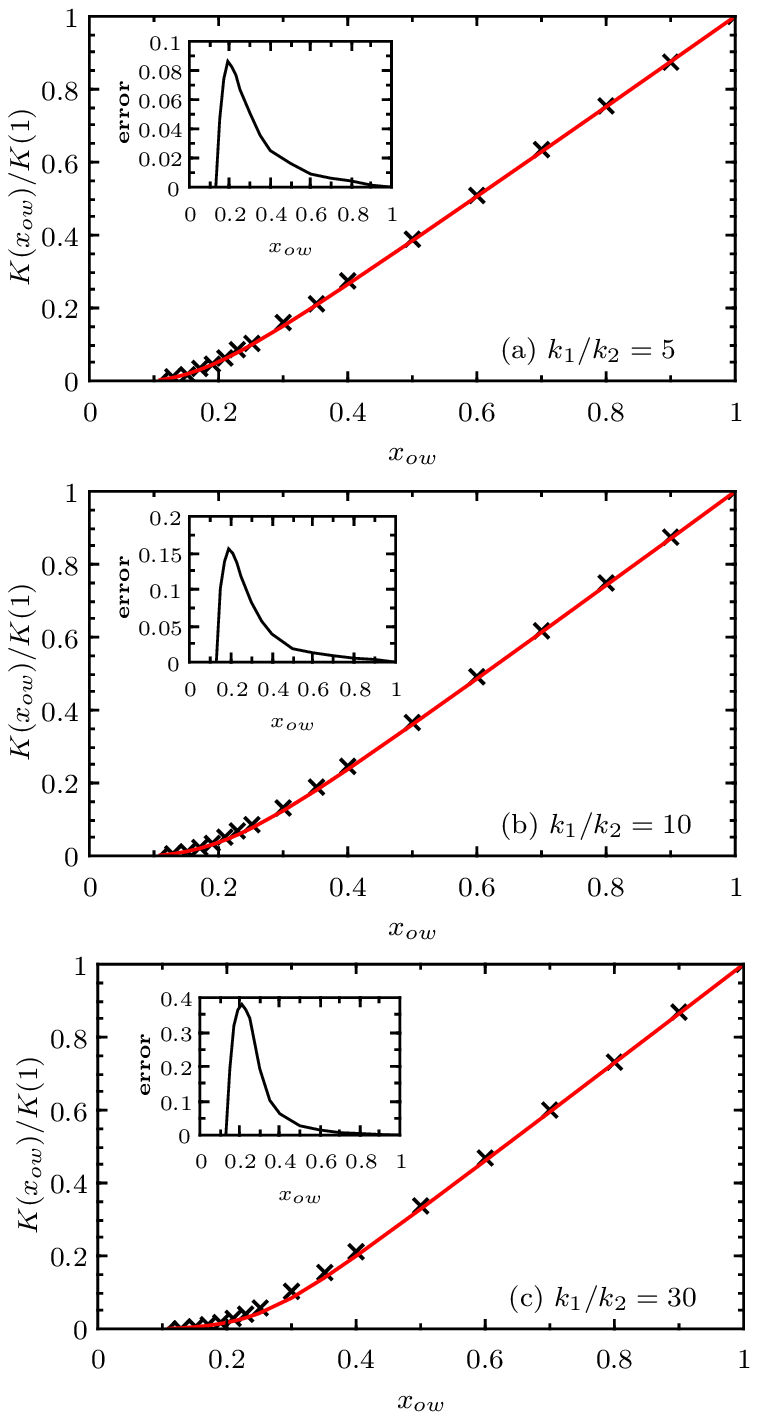}

\caption{\label{fig:mix_model2}Comparison of ``EM1'' model predictions (red
line) to network simulations (black cross) with $N_{v}=14$. Insets
are the relative error of the EM model prediction.}
\end{figure}

(iii) This configuration corresponds to the latter case but with two
local permeabilities having the same fraction $x_{1}=x_{2}$, so that
Eqs. \ref{eq:mean_local_perm1} and \ref{eq:effec_perm1} can be used.
Comparison with pore-network simulations is shown in Fig. \ref{fig:mix_model2}
for various ratios of local permeabilities $k_{2}/k_{1}$ with $N_{v}=14$.
As already mentionned, EM model reproduces correctly results from
pore-network simulations (fig. \ref{fig:mix_FEM}). However, for low
ratio $k_{2}/k_{1}$, deviations are observed in the vicinity of the
percolation threshold.

\section{Comparison with experiments}

In the next section, we detail experiments conducted on real solid
foams: foaming process, microstructural characterization, permeability
measurements, etc. Finally, we compare experimental results with predictions
of EM model.

\subsection{Elaboration of controlled polymer foams}

\begin{figure}
\includegraphics[scale=0.9]{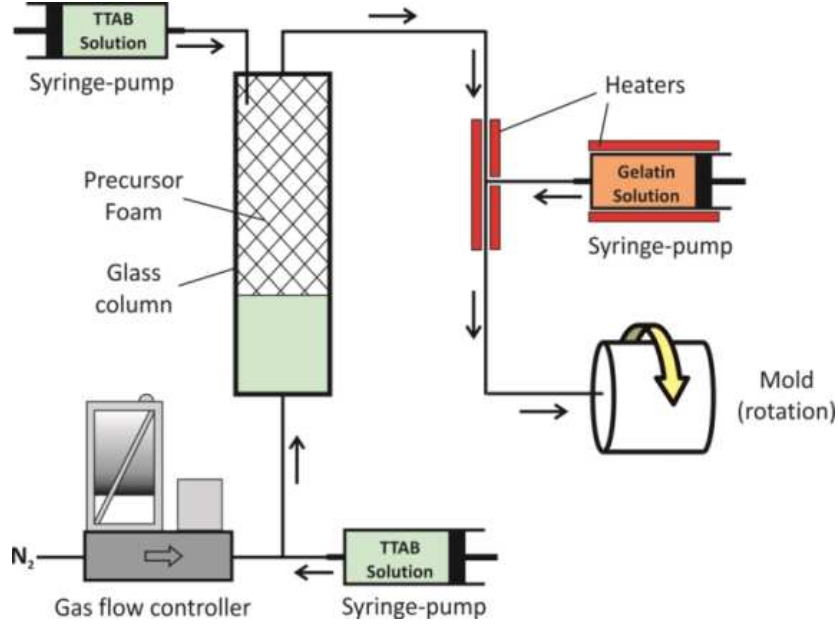}

\caption{\label{fig:Exp-setup}Foaming process}
\end{figure}

We elaborate solid polymer foam samples having fixed values for both
gas volume fraction and monodisperse bubble diameter $D_{b}$, but
a tunable membrane content. The experimental procedure can be described
as follows (see Fig. \ref{fig:Exp-setup}): (1) monodisperse precursor
aqueous foam is generated. Foaming liquid, i.e. TTAB (TetradecylTrimethylAmonium
Bromide) at 3 g/L in water, and nitrogen are pushed through a T-junction
allowing the bubble size control by adjusting the flow rate of each
fluid. Produced bubbles are collected in a glass column and a constant
gas fraction over the foam column is set at 0.99 by imbibition from
the top with foaming solution \citep{Lorenceau09}. (2) An aqueous
gelatin solution is prepared at a mass concentration $C_{gel}$ within
the range 12-18\%. The temperature of this solution is maintained
at $T$ \ensuremath{\approx} 60\textdegree C in order to remain above
the sol/gel transition ($T_{(s/g)}\approx$ 30\textdegree C). (3)
The precursor foam and the hot gelatin solution are mixed in a continuous
process thanks to a mixing device based on flow-focusing method \citep{khidas15,khidas14}.
By tuning the flow rates of both the foam and the solution during
the mixing step, the gas volume fraction can be set, $\phi_{0}$ =
0.8. Note also that the bubble size is conserved during the mixing
step. The resulting foamy gelatin is continuously poured into a cylindrical
cell (diameter: 40 mm and height: 40 mm) which is rotating around
its axis of symmetry at approximately 50 rpm. This process allows
for gravity effects to be compensated until the temperature decreases
below $T_{(s/g)}$. (4) The cell is let one hour at 5\textdegree C,
then one week in a climatic chamber ($T$ = 20\textdegree C and RH
= 30\%). During this stage, water evaporates from the samples and
the gas volume fraction increases significantly. (5) After unmolding,
a slice (thickness: 20 mm and diameter: 40 mm) is cut.

\subsection{Characterization of the foam samples}

\subsubsection*{Pore volume fraction:}

As the density of dried gelatin was measured to be 1.36, volume and
weight measurements of the dried foam samples give the pore volume
fraction. For the gelatin concentrations used in this study, the pore
volume fraction is found to vary between 0.977 and 0.983, so that
in the following we will consider that this parameter is approximately
constant and equal to 0.980\textpm 0.003.

\subsubsection*{Pore size:}

\begin{figure}
\includegraphics[scale=1.1]{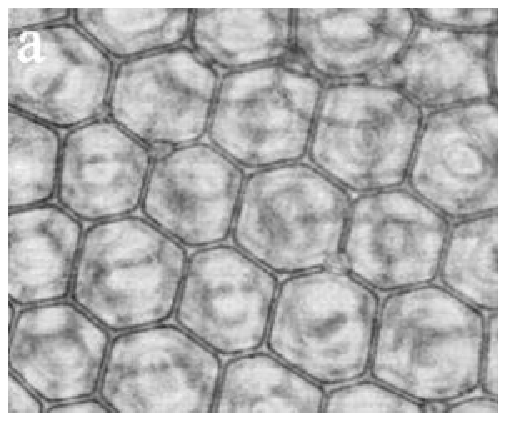}\includegraphics[scale=0.9]{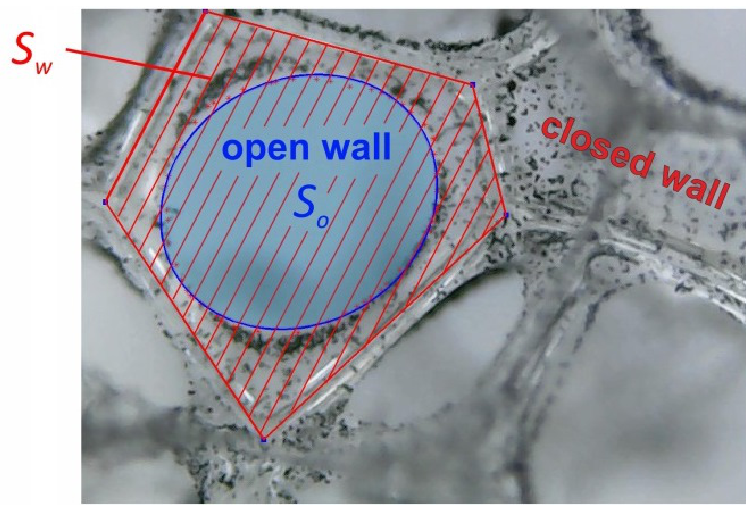}

\caption{\label{fig:foam}Characterization of foam samples}
\end{figure}

Through a preliminary calibration, observation of the sample surface
(see Fig. \ref{fig:foam}a) allows for the pore (bubble) size to be
measured. The calibration procedure can be described as follows: bubbles
collected in the glass column (precursor in the Fig. \ref{fig:Exp-setup})
are sampled and squeezed between two glass plates separated from 100
\textmu m. Then the surface exposed with a microscope is measured,
and using volume conservation, bubble gas volume is determined and
the mean bubble diameter $D_{b}$ is obtained with a precision better
than 3\%. Moreover, we measure the mean length $L_{p}$ that characterized
the Plateau borders of the precursor foam at the column wall. We obtain
the following relationship, $D_{b}=(1.68\pm0.06)L_{p}$, that can
be used afterwards for measuring the pore size in the dried gelatin
samples. We measure $D_{b}$ = 810 \textmu m (the absolute error on
$D_{b}$ is \textpm{} 30 \textmu m) for all the samples.

\subsubsection*{Cell wall characteristics:}

We characterize cell walls in our samples through observation with
a microscope on both top and bottom sample surfaces (see Fig. \ref{fig:foam}).
For each sample, a large number of cell walls, $N_{w}$>100, were
observed in order to determine the following parameters: the number
of open walls, $N_{ow}$, and the proportion of open walls:

\begin{equation}
x_{ow}=N_{ow}/N_{w}
\end{equation}

The proportion of closed walls is therefore $x_{cl}=1-x_{ow}$ . Note
that those open walls exhibit various aperture sizes, so the aperture
ratio of walls is measured:

\begin{equation}
\delta_{ow}^{\prime\prime}=(S_{o}\text{\textfractionsolidus}S_{w})^{1/2}
\end{equation}

where $S_{o}$ is the aperture area and $S_{w}$ is the total area
of the window (see Fig. \ref{fig:foam}).

\begin{figure}
\includegraphics[scale=1.0]{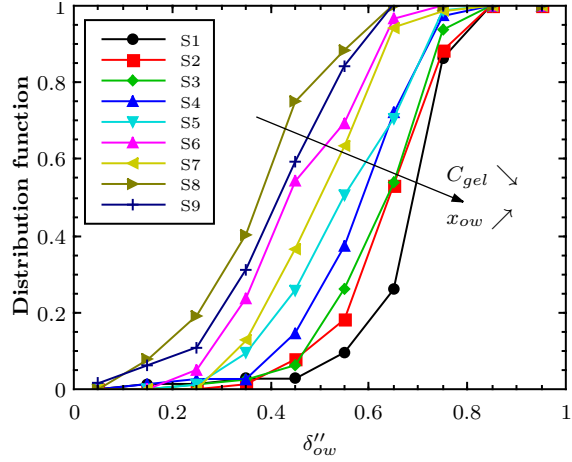}

\caption{\label{fig:distrib_function}Distribution function of aperture factors
$\delta_{ow}^{\prime\prime}$}
\end{figure}

The structural characterization is completed by a measurement of the
membrane thickness through SEM images. From nine micrographs the average
thickness has been measured to be equal to 1.5\textpm 0.25 \textmu m,
which is close to thicknesses measured for similar polymer foams \citep{yasunaga96,hoang14,gao16}.

Fig. \ref{fig:distrib_function} shows the distribution function of
aperture factor for the samples, and Table \ref{tab:Characteristics-of-foam}
gives their corresponding mean value $<\delta_{ow}^{\prime\prime}>$
and their proportion of open walls $x_{ow}$.

\begin{table}
\caption{\label{tab:Characteristics-of-foam}Characteristics of foam samples}

\begin{tabular}{|c|c|c|c|c|c|c|c|c|c|c|}
\hline 
\multicolumn{2}{|c|}{Sample} & S1 & S2 & S3 & S4 & S5 & S6 & S7 & S8 & S9\tabularnewline
\hline 
\hline 
\multicolumn{2}{|c|}{$C_{gel}$ (\%)} & 12 & 13 & 16 & 16 & 16 & 17 & 18 & 18 & 18\tabularnewline
\hline 
\multicolumn{2}{|c|}{$<\delta_{ow}^{\prime\prime}>$} & 0.72 & 0.68 & 0.67 & 0.62 & 0.59 & 0.50 & 0.54 & 0.42 & 0.46\tabularnewline
\hline 
\multicolumn{2}{|c|}{$x_{ow}$} & 0.93  & 0.83 & 0.79 & 0.69 & 0.60 & 0.54 & 0.32 & 0.22 & 0.15\tabularnewline
\hline 
\multirow{2}{*}{$10^{3}*K/D_{b}^{2}$ } & direct meas. & 17.63 & 9.10 & 6.99 & 3.83 & \multicolumn{5}{c|}{out of range}\tabularnewline
\cline{2-11} 
 & acoustic meas. & 16.89 & 8.02 & 7.58 & 2.79 & 2.33 & 1.43 & 0.579 & 0.140 & 0.017\tabularnewline
\hline 
\end{tabular}
\end{table}

\subsection{Foam permeability}

\subsubsection*{Permeability measurement:}

We determined the permeability by acoustic measurements performed
in a three-microphone impedance tube \citep{utsuno89,salissou10}.
Permeability value is deduced from the imaginary part of the low frequency
behavior of the effective density \citep{panneton06}: $K=-\mu/\lim_{\omega\to0}[\Im(\omega\rho)]$.
Note that the diameter of the samples is slightly larger than 40 mm
so that air leakage issue and sample vibration were successfully avoided.
The air permeability is determined on frequencies ranging from 200
Hz to 400 Hz.

Samples showing high permeability, i.e. $K>10^{-9}m^{2}$, were characterized
by a direct measurement of the pressure drop $\Delta P_{sp}$ as a
function of air flow rate $Q$ within steady laminar conditions, and
the Darcy permeability was determined as follows\citep{airflow}:

\begin{equation}
K=\mu QH/A\Delta P_{sp}
\end{equation}

with the thickness of sample $H$\ensuremath{\approx} 20 mm and the
circular cross-sectional area $A$\ensuremath{\approx}1.25 $cm^{2}$.

\subsubsection*{Comparison to theoretical predictions:}

Theoretical calculations are performed by assuming the Sampson local
permeability ($k_{i}=d_{i}^{3}/24D_{b}$) in Eq. \ref{eq:mean_local_perm1}
and by using Eq. \ref{eq:effec_perm1}. The size of aperture $d_{i}$
is calculated from the aperture factor $\delta_{ow}^{\prime\prime}$
and the mean size of bubbles $D_{b}$:

\[
d_{i}=\delta_{ow}^{\prime\prime}\left\langle t_{w}/D_{b}\right\rangle D_{b}
\]

\begin{figure}[h]
\includegraphics[scale=1.0]{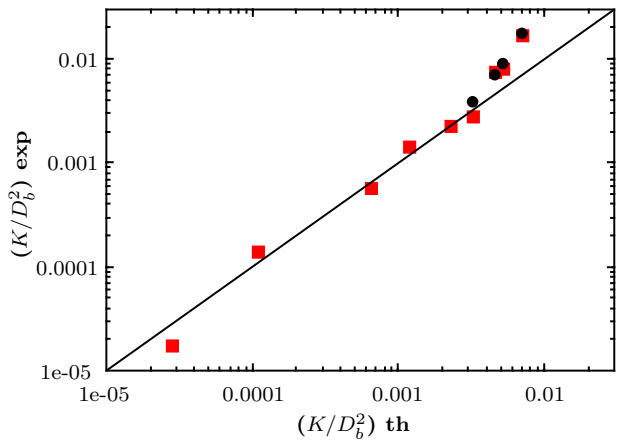}\caption{\label{fig:exp_vs_th}Comparison of direct (dot) and acoustical (square)
measurements of foam permeability to theoretical predictions. }
\end{figure}

The ratio $\left\langle t_{w}/D_{b}\right\rangle $ as well as the
pore neighbor number are assumed to be given by Kelvin cell structure:
$(\frac{8}{14}\sqrt{3}L+\frac{6}{14}L)/2\sqrt{2}L\approx0.5$4 and
$N_{v}=14$. Moreover, aperture distributions shown in Fig.\ref{fig:distrib_function}
allows to calculate the fraction $x_{i}$ of walls having an aperture
size equal to $d_{i}$. Fig. \ref{fig:exp_vs_th} shows that the theoretical
predictions are in good agreement with experimental measurements.
Various reasons can be mentioned to explain the discrepancy observed
at high permeability: error caused by assuming the real foam structure
like a Kelvin structure, error in measurement of aperture size (a
theoritical calculation of permeability with walls apertures 30\%
greater is enough to delete the gap), error in permeability measurement
due to air leak around the sample.

\section{Conclusion}

We have studied the static permeability of solid foams by combining
different approaches: (i) Numerical approach: FEM simulations computing
Stokes problem both at the pore scale and at the macro-scale, a simulation
of simplified flow performed on a network of interconnected pores
interacting by Sampson local permeabilities, (ii) Theoretical approach:
a model of effective permeability based on the same theoretical framework
than network simulations has been developed, (iii) Experimental approach:
controlled samples of gelatin foams have been prepared, dried and
fully characterized (bubble size, wall aperture, air permeability).
FEM simulations allowed us to validate the use of network simulations
to predict the effects on permeability of both the wall aperture size
and the amount of closed walls. Compared to FEM simulations, networks
simulations are low computational cost and can be widely used to test
the predictions of the EM model, and to study the percolation effect
of the pore network. These different approaches made it possible to
develop and validate a model of effective foam permeability (Eqs.
\ref{eq:local_sampson}, \ref{eq:x_op}, \ref{eq:Rop}, \ref{eq:mean_local_perm1}
and \ref{eq:effec_perm1}).

For future studies, network simulations could be useful to study permeability
of various porous foamy materials (topologically disordered foams
or containing double porosity,...) and to check the ability of EM
model to predict permeability of such materials.
\begin{acknowledgments}
This work was part of a project supported by ANRT (Grant no. ANR-13-RMNP-0003-01). 
\end{acknowledgments}

\section*{Appendix}

\begin{figure}[h]
\includegraphics[scale=0.5]{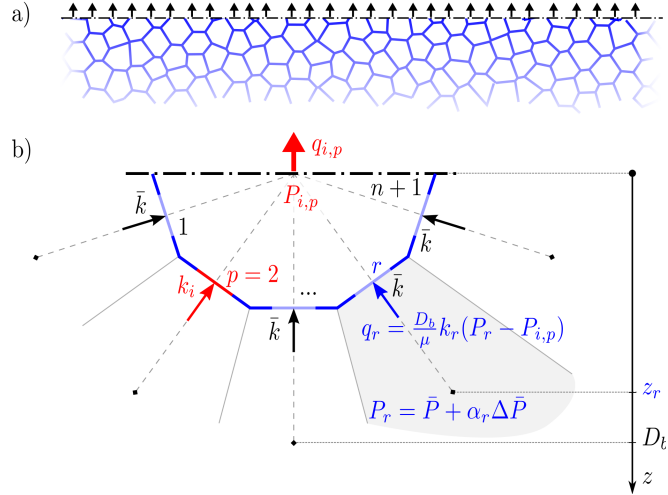}\caption{\label{fig:model_fig}(a) Cross-section of foam, (b) Geometry of a
half pore representative of pores contained inside the foam cross-section.
Note that we have to consider $n+1$ configurations for the position
$p$ of the membrane associated with the permeability $k_{i}$. Figure
depicts the case $p=2.$}
\end{figure}

Here, we detail the calculation of the mean local permeability. We
consider a cross-section of foam (fig. \ref{fig:model_fig}a) and
calculate the mean local permeability $\bar{k}$ of a foam containing
different local permeabilities $\left\{ k_{i}\right\} $. To represent
a pore inside the cross-section, we consider a half pore connected
to $N_{v}/2$ effective pores such as $n=\frac{N_{v}}{2}-1$ membranes
have a local permeability equal to the mean local permeability $\bar{k}$
and the last one located at the $p^{th}$ position has a permeability
equal to $k_{i}$ (fig. \ref{fig:model_fig}b). Due to the heterogeneity
induced by the local permeability $k_{i}$, the pressure inside the
central pore $P_{i,p}$ is different from the mean pressure $\bar{P}$.
Pressures inside neighbor effective pores are supposed equal to the
effective pressure expected for each peculiar position of the neighbor
pore: $\bar{P}+\alpha_{r}\Delta\bar{P}$ with $\alpha_{r}=z_{r}/D_{b}$.
The total flow rate passing through the central half pore is equal
to:

$q_{i,p}=\frac{D_{b}}{\mu}\sum_{r=1}^{n+1}q_{r}=\frac{D_{b}}{\mu}\sum_{r=1}^{n+1}k_{r}\left(\bar{P}+\alpha_{r}\Delta\bar{P}-P_{i,p}\right)$

where $k_{r}=\bar{k}$ for $r\neq p$, and $k_{i}$ for $r=p$.

The total flow rate can be written in a more useful way as:

$q_{i,p}=\frac{D_{b}}{\mu}\left[\sum_{r=1}^{n+1}\bar{k}\left(\bar{P}-P_{i,p}+\alpha_{r}\Delta\bar{P}\right)+\left(k_{i}-\bar{k}\right)\left(\bar{P}-P_{i,p}+\alpha_{p}\Delta\bar{P}\right)\right]$
.

The effective flow rate $\bar{q}$ passing through the effective pore
is obtained by considering $k_{i}=\bar{k}$ and $P_{i,p}=\bar{P}$
in the previous equation:

$\bar{q}=\frac{D_{b}}{\mu}\sum_{r=1}^{n+1}\bar{k}\alpha_{r}\Delta\bar{P}$.

Therefore, the flow rate $q_{i,p}$ can be expressed in function of
$\bar{q}$:

$q_{i,p}=\bar{q}+\frac{D_{b}}{\mu}\left[\left(n+1\right)\bar{k}\left(\bar{P}-P_{i,p}\right)+\left(k_{i}-\bar{k}\right)\left(\bar{P}-P_{i,p}+\alpha_{p}\Delta\bar{P}\right)\right]$.

In the following, we suppose that the total flow rate passing through
the central half pore $q_{i,p}$ is equal to the effective flow rate
$\bar{q}$ leading to: 

$0=\left(n+1\right)\bar{k}\left(\bar{P}-P_{i,p}\right)+\left(k_{i}-\bar{k}\right)\left(\bar{P}-P_{i,p}+\alpha_{p}\Delta\bar{P}\right)$.

This hypothesis leads to the pressure inside the central pore:

\[
P_{i,p}=\bar{P}+\frac{\left(k_{i}-\bar{k}\right)}{n\bar{k}+k_{i}}\alpha_{p}\Delta\bar{P}
\]

Now, we may impose the self-consistency condition, requiring that
the average $\left\langle P_{i,p}\right\rangle _{p,i}=\left\langle \left\langle P_{i,p}\right\rangle _{p}\right\rangle _{i}$
is equal to the effective pressure $\bar{P}$ leading to:

\[
\left\langle \frac{k_{i}-\bar{k}}{n\bar{k}+k_{i}}\right\rangle _{i}\left\langle \alpha_{p}\right\rangle _{p}\Delta\bar{P}=0
\]

The previous equation can be rewritten in an alternative form:
\[
\left\langle \frac{1}{n\bar{k}+k_{i}}\right\rangle _{i}=\frac{1}{\left(n+1\right)\bar{k}}
\]

To determine the macroscopic effective permeability, we calculate
the macroscopic flow rate $Q$ passing through the whole cross-section
$A$ containing $N_{w}$ walls having a local permeability equal to
$\bar{k}.$ Moreover, we suppose that the effective gradient of pressure
around the cross-section $\frac{\Delta\bar{P}}{D_{b}}$ is equal to
the mean pressure gradient $\frac{\Delta P_{sp}}{H}$. Then the macroscopic
flow rate is given by:

\[
Q=\frac{D_{b}}{\mu}\left[\sum_{w=1}^{N_{w}}\alpha_{w}\right]\bar{k}\Delta\bar{P}=\frac{D_{b}}{\mu}*N_{w}\left\langle \alpha\right\rangle _{p}*\bar{k}*\frac{\Delta P_{sp}}{H}D_{b}
\]

leading to the macroscopic effective permeability:

\[
K=\frac{N_{w}}{A}\left\langle \alpha\right\rangle _{p}D_{b}^{2}\bar{k}
\]

In considering the continuous limit for the calculation of $\left\langle \alpha\right\rangle _{p}$,
we obtain: $\left\langle \alpha\right\rangle _{p}=\frac{1}{2\pi}\int_{0}^{2\pi}\int_{0}^{\pi/2}sin(\theta)cos(\theta)d\theta d\varphi=\frac{1}{2}$.
And in the case of a Kelvin structure, the surface wall density $N_{w}/A$
is equal to $n/D_{b}^{2}$.

In the case of a binary mixture of local permeabilities (e.g. fully
open foam), the mean local permeability $\bar{k}$ is given by the
following equation:

$\frac{\bar{k}}{k_{\infty}}=\frac{1}{2}\left[\alpha+\left(\alpha^{2}+4\left(1-\alpha\right)\frac{k_{0}}{k_{\infty}}\right)^{0.5}\right]$

with

$\alpha=1-\frac{k_{1}k_{2}}{nk_{0}k_{\infty}}$ 

$k_{\infty}=k_{Voigt}=x_{1}k_{1}+x_{2}k_{2}$

$k_{0}=k_{Reuss}=\left(\frac{x_{1}}{k_{1}}+\frac{x_{2}}{k_{2}}\right)^{-1}$

$k_{\infty}$ and $k_{0}$ correspond respectively to the permeability
of an infinitely interconnected network ($N_{v}\rightarrow\infty$)
and this one of a poorly interconnected network ($N_{v}=2$).

\bibliographystyle{plainnat}

\end{document}